\documentclass[aps,pra,noeprint,longbibliography,reprint]{revtex4-1} 

\usepackage{graphicx}
\usepackage[hidelinks]{hyperref}
\usepackage{braket}

\usepackage[percent]{overpic} 
\usepackage{ulem} 
\usepackage{xcolor} 
\usepackage{mdframed} 
\newmdenv[backgroundcolor=green!10!white, linecolor=white, leftmargin=10pt, innerleftmargin=5pt,innertopmargin=5pt,innerrightmargin=5pt,innerbottommargin=5pt]{answer}

\newcommand{\unit}[1]{\,\mathrm{#1}}
\renewcommand{\vec}[1]{\mathrm{\mathbf{#1}}}

\newcommand{\textold}[1]{\textcolor{red!90!black}{\sout{#1}}}

\renewcommand{\textold}[1]{}

\begin{document}
    
    \title{A fermionic impurity in a dipolar quantum droplet}
    
    
    \author{Matthias Wenzel}
    \author{Tilman Pfau}
    \email{t.pfau@physik.uni-stuttgart.de}
    \author{Igor Ferrier-Barbut}
    
    \affiliation{5{.} Physikalisches Institut and Center for Integrated Quantum Science and Technology, Universit{\"a}t Stuttgart, Pfaffenwaldring 57, 70569 Stuttgart, Germany}

    \date{\today}
    
    \begin{abstract}
        In this article we develop the framework to describe Bose-Fermi mixtures of magnetic atoms, focusing on the interaction of bosonic self-bound dipolar quantum droplets with a small number of fermions. We find an attractive interaction potential due to the dipolar interaction with several bound states, which can be occupied by one fermion each, resulting in a very weak back-action on the bosons. We conclude, that these impurities might act as unique probes giving access to inherent properties of dipolar quantum droplets.
    \end{abstract}
    
    
    
    \maketitle
    
    Self-bound quantum droplets \cite{Petrov2015,Schmitt2016,Cabrera2018,Semeghini2018} are manifestations of exotic liquid-like states of matter. They are reminiscent of superfluid helium droplets, but exist at orders of magnitude lower densities. Dipolar quantum droplets arise in Bose-Einstein condensates of strongly magnetic atoms. A combined experimental \cite{Kadau2016,Ferrier-Barbut2016,Schmitt2016,Chomaz2016a,Wenzel2017} and theoretical \cite{Petrov2015,Wachtler2016,Bisset2016,Wachtler2016a,Baillie2016,Baillie2017,Saito2016,Macia2016,Cinti2017} effort has recently revealed, that these droplets are stabilized by beyond-mean-field effects, but many questions about their dynamics and finite temperature behavior still remain. 
    In the case of helium droplets, remarkable insights and applications have been made possible by the immersion of impurities \cite{Toennies2004}. In this spirit, we investigate the interaction of magnetic fermionic dipolar impurities with a bosonic dipolar quantum droplet. 
    Immersed in the droplet, these impurities are subject to an intrinsically attractive dipole-dipole interaction.
    Therefore, we study the behavior of Bose-Fermi mixtures with dominant dipole-dipole interaction (DDI). In the limit of only a few fermions, we show that the Hamiltonian reduces to a simple Schroedinger equation with the interspecies interaction acting as a trapping potential for the fermions. Solving it, we find several anharmonically bound states. Due to the Pauli exclusion principle and the shallowness of the attractive potential we find that only a few fermions can be trapped. Thus, the few-fermion limit is intrinsically satisfied.

    \begin{figure}
    	\begin{overpic}[width=0.48\textwidth]{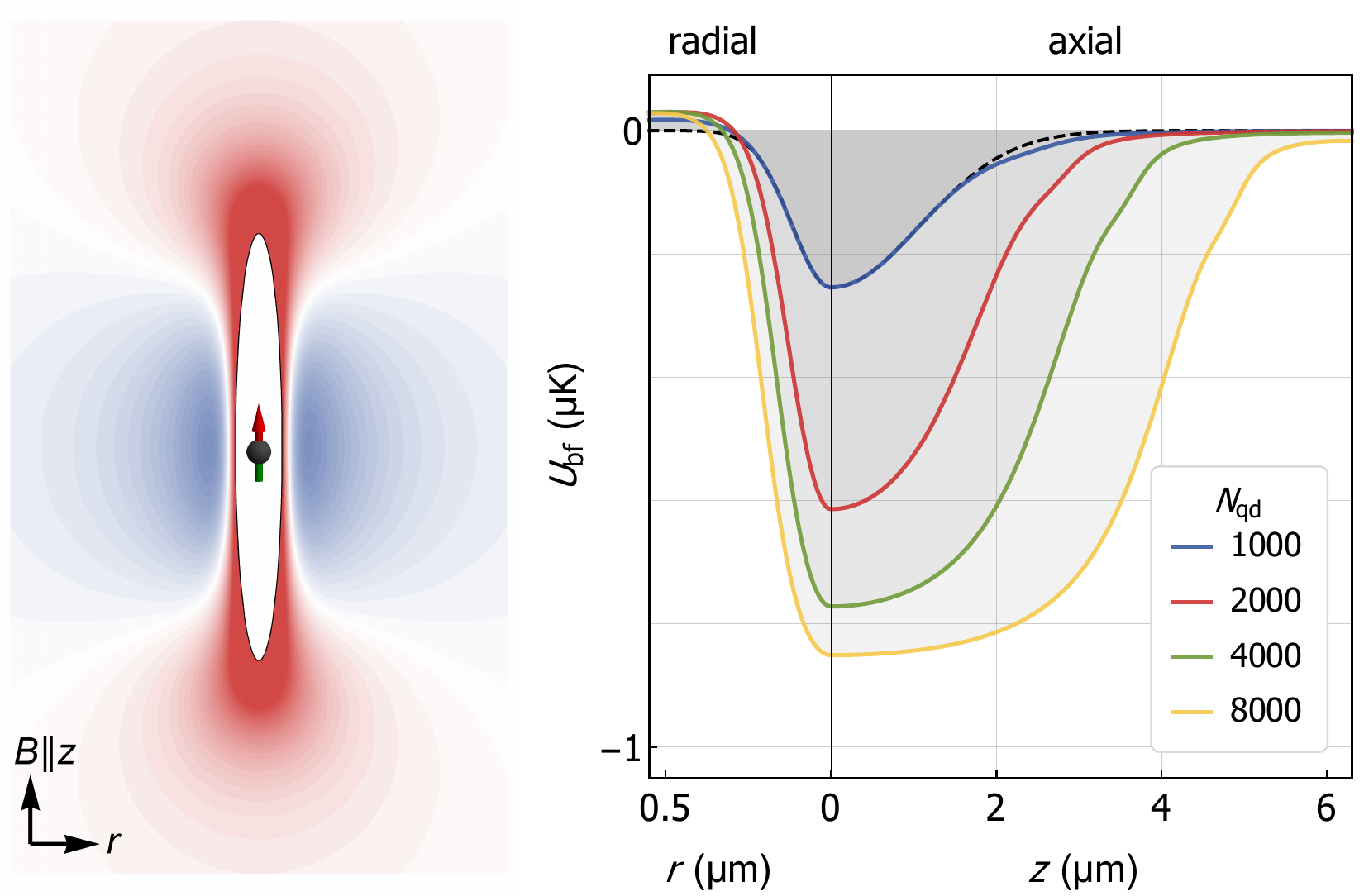}
    		\put(2,60){a)} \put(41,60){b)}\end{overpic}
        \vspace{-5mm}
        \caption{\label{fig:1} a) Schematic of a dipolar fermionic impurity interacting with bosons in a dipolar quantum droplet (white ellipse). The interaction potential is attractive (red) axially and repulsive radially (blue). 
        b) Calculated mean-field trapping potentials $U_\mathrm{bf}(r,z)$ according to eq.~(\ref{eq:Ubf}) for a interspecies scattering length $a_\mathrm{bf} = 70\,a_0$, $a_0$ being the Bohr radius, and different atom numbers $N_\mathrm{qd}$ of the droplet. The dipolar interaction leads to deviations from a Gaussian density profile (dashed).}
    \end{figure}
    
    Starting within second quantization, the full Hamiltonian of the system $\hat{H} = \hat{H}_\mathrm{f} + \hat{H}_\mathrm{b} + \hat{H}_\mathrm{bf}$ in its most general form is composed of the intraspecies contribution
    \begin{eqnarray}\label{eq:Hf}
        \hat{H}_\mathrm{f} &=& \frac{\hbar^2}{2m_\mathrm{f}} \int\!\!\mathrm{d}\vec{r} \, \nabla\hat{\psi}^\dagger_\mathrm{f} \cdot \nabla\hat{\psi}_\mathrm{f} \nonumber\\*
        &+& \frac{1}{2} \int\!\!\mathrm{d}\vec{r}\,\mathrm{d}\vec{r^\prime}\, \hat{\psi}^\dagger_\mathrm{f}(\vec{r^\prime}) \hat{\psi}^\dagger_\mathrm{f}(\vec{r}) \, U_\mathrm{ff}(\vec{r}-\vec{r^\prime}) \,\hat{\psi}_\mathrm{f}(\vec{r}) \hat{\psi}_\mathrm{f}(\vec{r^\prime}) \qquad
    \end{eqnarray}
    for fermions and similarly for bosons, and the interspecies interaction
    \begin{equation}\label{eq:Hbf}
        \hat{H}_\mathrm{bf} = \frac{1}{2} \int\!\!\mathrm{d}\vec{r}\,\mathrm{d}\vec{r^\prime}\, \hat{\psi}^\dagger_\mathrm{f}(\vec{r^\prime}) \hat{\psi}^\dagger_\mathrm{b}(\vec{r}) \, U_\mathrm{bf}(\vec{r}-\vec{r^\prime}) \,\hat{\psi}_\mathrm{b}(\vec{r}) \hat{\psi}_\mathrm{f}(\vec{r^\prime}) 
        \,. \quad
    \end{equation}
    The predominant interactions for ultra-cold dipolar atoms are the contact interaction, which can be written in pseudo-potential form $U_\mathrm{con}(\vec{r}) = g\,\delta(\vec{r}) \frac{\partial}{\partial |\vec{r}|}|\vec{r}|$ with $g = 4\pi\hbar^2a/m$ defined by the s-wave scattering length $a$ and the atomic mass $m$, as well as the DDI
    \begin{equation}
        U_\mathrm{dip}(\vec{r}) = \frac{\mu_0 \mu^2}{4\pi} \frac{1-3 \cos(\theta)^2}{|\vec{r}|^3} \,,
    \end{equation}
    which is non-local due to its long-range character and defined by the magnetic moment $\mu$ of the atom \cite{Lahaye2009}. There is no contact interaction between fermions in identical spin states, but they are subject to the dipolar interaction \cite{Lu2012}.
    
    We are particularly interested in the ground state of few fermions, where the number of fermions $N_\mathrm{f}$ is small compared to the number of bosons $N_\mathrm{b} \gg N_\mathrm{f}$. This restriction leads to a few assumptions.
    First, we neglect any back-action of fermions on the bosonic many-body state. Second, scaling with atom number, the interspecies interaction is much stronger than the fermionic intraspecies DDI.  Third, the bosons are subject to quantum depletion \cite{Schutzhold2006}, which is small even for systems that are stabilized by quantum fluctuations \cite{Wachtler2016a,Baillie2016}.
    Therefore we neglect any interaction of depleted bosons with fermions and the DDI between fermions. For both effects we estimate the order of magnitude later.

    With these assumptions, the bosonic many-body state is not modified by the presence of fermions. The former is then described by the extended Gross-Pitaevskii equation (eGPE) \cite{Wachtler2016,Bisset2016,Wachtler2016a,Baillie2016,Saito2016} within the first beyond-mean-field correction \cite{Lima2011}, yielding the bosonic density distribution $n_\mathrm{b}(\vec{r})$ and ground state energy $E_\mathrm{b}$.
    For the interspecies interaction $\hat{H}_\mathrm{bf}$, we consequently replace the bosonic operator $\hat{\psi}_\mathrm{b}(\vec{r})$ by the wavefunction $\psi_\mathrm{b} = \sqrt{n_\mathrm{b}(\vec{r})}$, which yields the Hamiltonian
    \begin{equation}\label{eq:H2}
        \hat{H} = E_\mathrm{b} + \int\!\!\mathrm{d}\vec{r}
        \left[ \frac{\hbar^2}{2m_\mathrm{f}} \nabla\hat{\psi}^\dagger_\mathrm{f} \cdot \nabla\hat{\psi}_\mathrm{f} + \frac{1}{2}
        \hat{\psi}^\dagger_\mathrm{f} \hat{\psi}_\mathrm{f} \, U_\mathrm{bf}(\vec{r})
        \right]
    \end{equation}
    with the Bose-Fermi interaction potential
    \begin{equation}\label{eq:Ubf}
            U_\mathrm{bf}(\vec{r}) = g_\mathrm{bf}\, n_\mathrm{b}(\vec{r}) + \int\!\!\mathrm{d}\mathbf{r}^\prime\, U_\mathrm{dip}\!\left(\vec{r}-\vec{r^\prime}\right) n_\mathrm{b}(\vec{r^\prime})
    \end{equation}
    relying on the interspecies scattering length $a_\mathrm{bf}$ defining $g_\mathrm{bf} = 4\pi\hbar^2 a_\mathrm{bf}/m_\mathrm{f}$ \footnote{We assume equal masses $m_\mathrm{f} \approx m_\mathrm{b}$ and therefore use $m_\mathrm{f}$ instead of introducing the reduced mass.}.
    Finally, the remaining problem of eq.~(\ref{eq:H2}) thus reduces to solving the stationary Schroedinger equation
    \begin{equation}\label{eq:seq}
        H =  -\frac{\hbar^2 \Delta^2}{2m_\mathrm{f}} + U_\mathrm{bf}(\vec{r})
    \end{equation}
    of a single particle in an external potential $U_\mathrm{bf}(\vec{r})$.

    As mentioned in the introduction, we are interested in the behavior of fermionic dipolar impurities interacting with a self-bound dipolar quantum droplet consisting of many bosonic atoms.
    Therefore we consider a number $N_\mathrm{qd}$ of bosonic $^{164}$Dy atoms with intraspecies scattering length $a_\mathrm{bb} = 70\,a_0$ in accordance with recent measurements \cite{Ferrier-Barbut2018}. The magnetic moment $\mu = 9.93\,\mu_\mathrm{B}$, $\mu_\mathrm{B}$ being the Bohr magneton, relies on the electronic configuration and is thus constant across dysprosium isotopes. For these parameters, the critical atom number for the stability of a self-bound dipolar quantum droplet is $N_\mathrm{qd} > N_\mathrm{crit} \approx 975$, as calculated with the eGPE \cite{Schmitt2016}. This way we also extract its density profile $n_\mathrm{b}(\vec{r})$ having cylindrical symmetry with respect to the polarization axis $\hat{\vec{z}}$.
    In a next step, we calculate the interaction potential $U_\mathrm{bf}(r,z) = U_\mathrm{bf}(\vec{r})$ of eq.~(\ref{eq:Ubf}) acting as an external trapping potential for the fermionic impurities.
    The necessary interspecies scattering length $a_\mathrm{bf}$, describing the Bose-Fermi contact interaction, is not known for dysprosium. Yet, we expect to find a suitable combination of isotopes and magnetic field owing to the abundance of Feshbach resonances in lanthanide atoms \cite{Maier2015a}. For the purpose of this manuscript we consider $^{163}$Dy with $a_\mathrm{bf} = 70\,a_0$ for now.
    As sketched in Fig.~1a, the resulting interaction potential resembles the droplet density distribution having radial symmetry and additionally attractive wings axially (red) and repulsive ones radially (blue) outside of the droplet (white ellipse). Furthermore, the trapping potential is highly anisotropic, as shown in Fig.~1b, with a strong dependence of the potential depth $U_0 = U_\mathrm{bf}(0)$ on the atom number $N_\mathrm{qd}$. 
    Owing to the incompressible character of the droplet, its peak density and thus $U_0$ saturate for $N_\mathrm{qd} \gtrsim 10^4$ atoms, while it grows axially in this regime \cite{Wachtler2016a}.

    \begin{figure}
    	\begin{overpic}[width=0.48\textwidth]{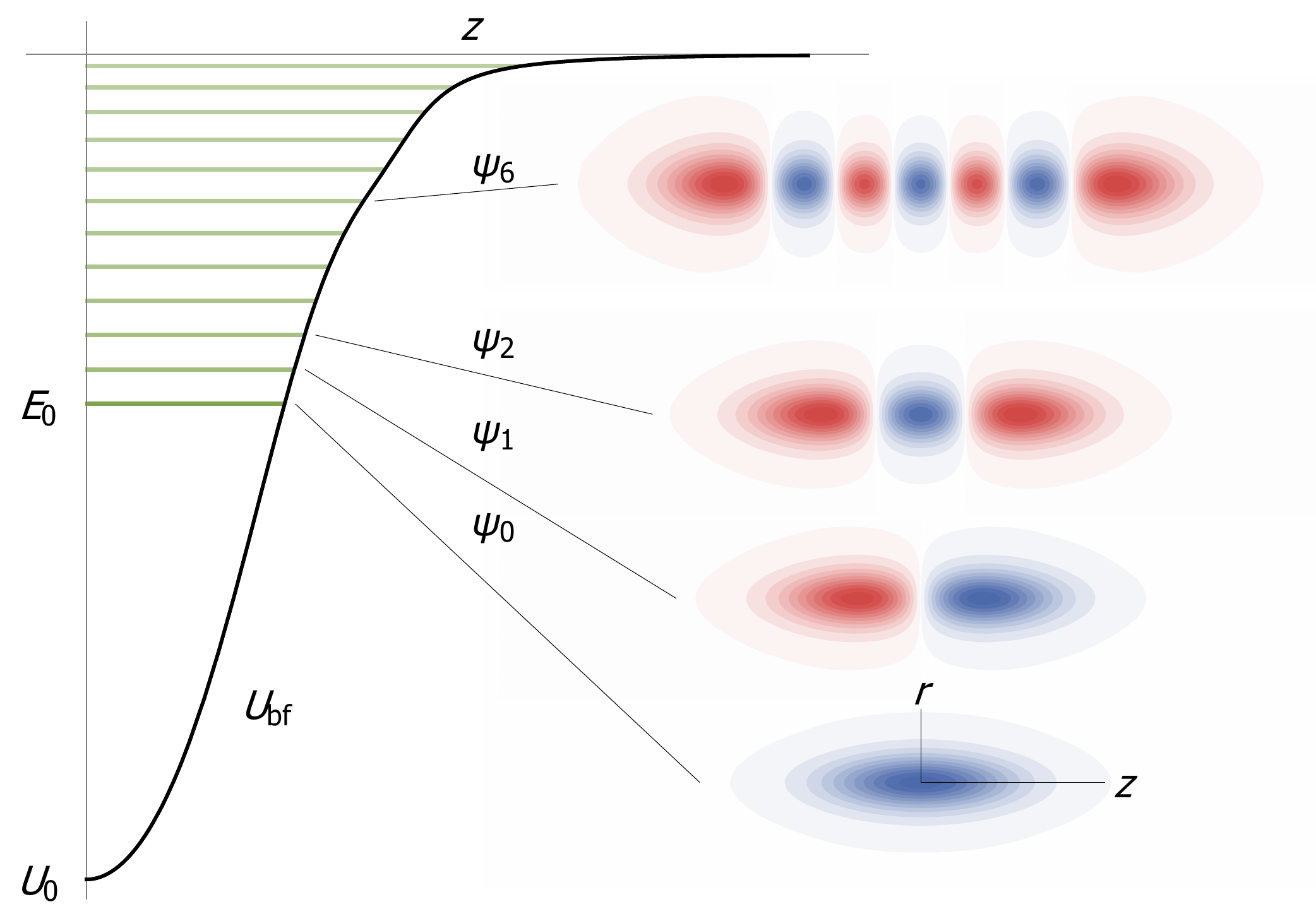}
    		\end{overpic}
        \vspace{-5mm}
        \caption{\label{fig:2} Bound states of an impurity for $N_\mathrm{qd} = 1500$ and $a_\mathrm{bf} = 70\,a_0$. The axial potential $U_\mathrm{bf}(r=0,z)$ is plotted (black) along with the bound state energy $E_\mathrm{s}$ (green). Insets show example wavefunctions $\psi_s(r,z)$, which resemble harmonic oscillator eigenstates. Having ground state character radially, these are one-dimensional systems. }
    \end{figure}
    
    In order to analyze the potential $U_\mathrm{bf}$ trapping the impurity, we Taylor-expand it at the center to obtain the trap frequencies $\omega_{r,z}$. Due to the anisotropy of the droplet the radial confinement $\omega_r \gg \omega_z$ is stronger than the axial one and comparable to the potential depth $U_\mathrm{0} \gtrsim \hbar \omega_r$ for the parameter range discussed in the following. Therefore we expect only a single bound state radially, while there are likely multiple axial states. Thus, this is effectively a one-dimensional system.
    
    In order to obtain the bound states we calculate the spectrum of eq.~(\ref{eq:seq}). Since the problem has cylindrical symmetry we expand on the basis set $\{\phi_{n l k}(r,\varphi,z)\}$ of the cylindrical harmonic oscillator, see appendix, and note that states with different $l$ quantum number are decoupled. Being in the radial ground state we restrict our analysis to the $l=0$ components.
    To simplify calculations, we approximate the potential $U_\mathrm{bf,sep}(r,z) \approx U_\mathrm{bf}(r,0)U_\mathrm{bf}(0,z)$, such that radial and axial contributions decouple.
    In the next step we compute the Hamiltonian matrix $\braket{H} = \braket{\phi_{n^\prime l^\prime k^\prime} | H | \phi_{n l k}}$ including radial (axial) quantum numbers $n = 0,2,...,24$ ($k = 0,1,...,59$) and $l=0$. The number of basis vectors is chosen such that all contributions $\ge 10^{-2}$ to the calculated eigenenergies are considered.  
    We obtain the latter by diagonalizing the Hamiltonian matrix and focus on bound state solutions $E_s < 0$ in the following. Fig.~2 shows an example with $N_\mathrm{qd} = 1500$ and $a_\mathrm{bf} = 70\,a_0$, where we find $N_\mathrm{bs} = 12$ bound states. A few corresponding eigenstates $\psi_s$ are depicted as well. Qualitatively, these are similar to the well-known harmonic oscillator states and have $\phi_{0 0 k}$ character. Thus the bound states are non-degenerate.

    \begin{figure}
    	\begin{overpic}[width=0.48\textwidth]{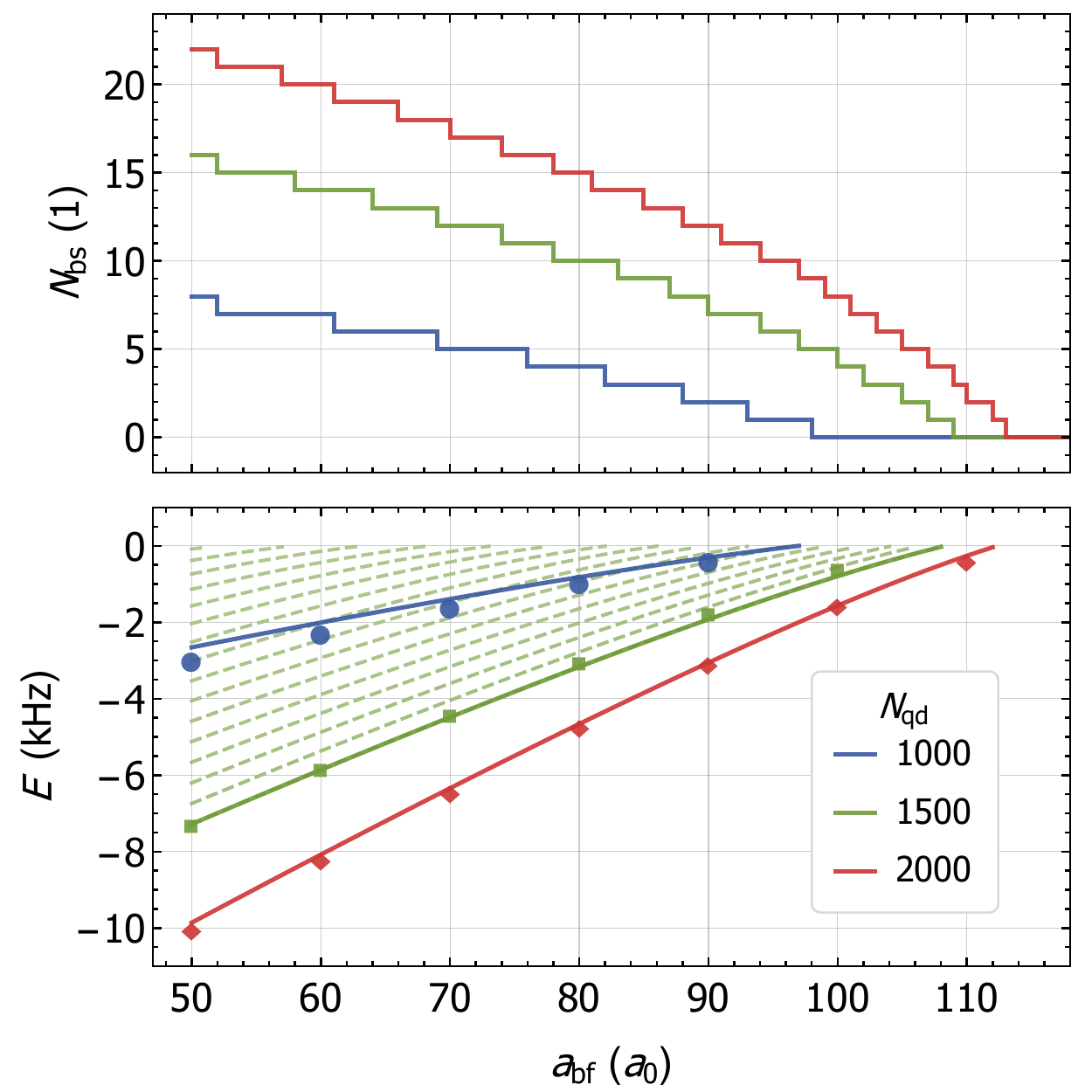}
    		\put(0,95){a)} \put(0,50){b)}\end{overpic}
        \vspace{-3mm}
        \caption{\label{fig:3} Bound state properties.
        a) Number of bound states $N_\mathrm{bs}$ over Bose-Fermi scattering length $a_\mathrm{bf}$ for various droplet atom numbers $N_\mathrm{qd}$.
        b) Ground state energy $E_0$ (solid lines) and corresponding calculations with the eGPE (dots). Additionally, excitated states $E_s> E_0$ are shown for $N_\mathrm{qd} = 1500$ (dashed).}
    \end{figure}
    
    In the following we focus on the parameter range of $N_\mathrm{qd} \le 2000$ atoms and $a_\mathrm{bf} = 50 - 120\,a_0$. In this regime, the potential depth is large enough to accomodate several bound states, while it is well in the one-dimensional regime.
    At the same time the number of bound states $N_\mathrm{bs}$ restricts the number of fermions in the system. Therefore the initial assumption $N_\mathrm{qd} \gg N_\mathrm{bs}$ neglecting an influence of fermions on the bosonic ground state is satisfied intrinsically.
    The number of bound states $N_\mathrm{bs}$ is a direct measure of the trap depth for a known droplet. Therefore it increases with increasing droplet atom number $N_\mathrm{qd}$ and decreasing Bose-Fermi scattering length $a_\mathrm{bf}$, as shown in Fig.~3a. The calculated energy $E_0$ of the ground state (solid lines) is verified by independent calculations via imaginary time evolution on a grid (dots), see Fig.~3b. For $N_\mathrm{qd} = 1500$ the spectrum $E_s$ of all bound states (dashed) is shown as well.
    The anharmonicity of the trapping potential $U_\mathrm{bf}$ leads to a rich level scheme with a typical spacing of $\Delta E = E_{s+1} - E_s \approx 500\unit{Hz}$, that decreases to $\approx 200\unit{Hz}$ towards the threshold.
    Experimentally, the level spectrum could be probed by driving transitions between bound states through harmonic modulation of $a_\mathrm{bf}$ at the frequency $\Delta E/h$.
    
    We now estimate the magnitude of the DDI between fermions, that we neglected so far.
    Although s-wave scattering is forbidden due to spin statistics, fermions can still interact via the dipolar interaction \cite{Lu2012}. To estimate this effect, we compute the Hartree energy
    \begin{equation}
        E_{s,s^\prime} = \frac{1}{2} \int\!\mathrm{d}\vec{r}\,\mathrm{d}\vec{r^\prime}
        \left|\psi_{s}(\vec{r})\right|^2 U_\mathrm{dip}\!\left(\vec{r}-\vec{r^\prime}\right) \left|\psi_{s^\prime}(\vec{r^\prime})\right|^2
    \end{equation}
    between the states $\psi_s$ and $\psi_{s^\prime}$, occupied with one fermion each. For the conditions of Fig.~2 the energy shift $E_{0,1}$ between ground and first excited state is $\approx 20\unit{Hz}$, which is small compared to the level spacing $\Delta E$, as mentioned earlier. For other combinations of $s$ and $s^\prime$ the values are even lower due to decreasing overlap of the wavefunctions. 
    Therefore, the initial assumption of negligible dipolar interaction between fermions appears satisfied.
    Full Hartree-Fock calculations could allow to calculate the modifications to the orbitals due to the DDI when several fermions are present. However, the interaction of fermions with the bosonic quantum depletion has to be taken into account as well.
    In order to estimate it, we calculate the condensate depletion $\Delta n / n \approx 5\,\%$ \cite{Lima2011} corresponding to $N_\mathrm{depl} = 75$ depleted atoms for a droplet with $N_\mathrm{qd} = 1500$. With $N_\mathrm{depl} > N_\mathrm{bs}$ and based on the prior estimate, this interaction can become sizeable. Thus, the energy due to interactions with the quantum depletion surpasses the fermion-fermion DDI interaction energy. Expanding on this idea, a fermionic impurity might be used to probe the quantum depletion. Yet, a more sophisticated theory is needed to describe and understand both effects properly, which is beyond the scope of this article.

    In conclusion, we derived the hamiltonian of fermions interacting with a large number of bosons. Leaving the bosonic state unaffected due to the low fermion number, the problem reduces to a Schroedinger equation with an external potential due to the Bose-Fermi interaction.
    We use this formalism to investigate the interaction of dipolar fermions with a dipolar self-bound quantum droplet. The resulting interaction potential closely resembles the elongated droplet density distribution with additional long-range features. For typical conditions of the droplet, the latter is attractive and we find a limited number of bound states. Since every state can only be occupied by a single fermion, we infer that neglecting a back-action on the bosonic droplet is justified, and the fermions act as impurities. These are effective one-dimensional systems due to the large aspect ratio of dipolar droplets.

    
    Finally, this is the first step towards probing quantum droplets with impurities. With negligible back-action on the droplet, this should be possible non-destructively on the bosons, using recently developed single atom detection techniques \cite{Bergschneider2018}.
    Up to now, the droplet temperature has not been measured, because time-of-flight expansion, the standard tool for thermometry, does not work with self-bound objects. Since the fermionic impurities thermalize with the bosonic environment, the presented excitation spectrum should be subject to thermal broadening. This way, we expect to extract the temperature and other finite-temperature effects, e.g. thermal fluctuations aiding to stabilize self-bound droplets, as recently suggested \cite{Boudjemaa2017,Aybar2018}.
    On the other hand, it is an open question whether self-evaporation towards zero temperature, that has been predicted for quantum droplets of Bose-Bose mixtures \cite{Petrov2015}, also occurs in its dipolar counterpart.
    
    \begin{acknowledgments}
        We thank the rest of the Dy team, Fabian Böttcher, Jan-Niklas Schmidt and Tim Langen, for proofreading of the manuscript. This work is supported by the German Research Foundation (DFG) within FOR2247 under Pf381/16-1, Pf381/20-1, and HBFG INST41/1056-1. IFB acknowledges support from the EU within Horizon2020 Marie Sk{\l}odowska Curie IF (703419 DipInQuantum).
    \end{acknowledgments}

    \appendix*

    \section{Cylindrical Harmonic Oscillator}
    For convenience, we show the solutions $\phi_{n l k}$ to the cylindrical harmonic oscillator, as used in the manuscript. The corresponding differential equation
    \begin{equation}
        \left[ -\frac{1}{2}\Delta + \frac{1}{2}(r^2 + z^2) - E \right] \!
        \psi = 0
    \end{equation}
    with $\Delta = \left(\partial_r^2 + \frac{1}{r}\partial_r + \frac{1}{r^2}\partial_\varphi^2 + \partial_z^2 \right)$ is separated into an axial and radial equation with the ansatz
    \begin{equation}
        \phi_{n l k}(r,\varphi,z) = R_{nl}(r,\varphi) \, Z_{k}(z)
    \end{equation}
    yielding the energy eigenvalues $E = \left(n + k + \frac{3}{2}\right)$.
    
    The radial equation is a two-dimensional harmonic oscillator in polar coordinates
    \begin{equation}
        \left[ -\frac{1}{2}\left(\partial_r^2 + \frac{1}{r}\partial_r + \frac{1}{r^2}\partial_\varphi^2 \right) + \frac{1}{2}r^2 - E_n \right] \!
        R_{nl}(r,\varphi) = 0
    \end{equation}
    and is solved by the radial function
    \begin{equation}
        R_{nl}(r,\varphi) = \sqrt{\frac{ (\frac{n-|l|}{2})! }{ \pi \, (\frac{n+|l|}{2})!}} \,
        e^{-r^2/2} \, e^{i l \varphi} \, r^{|l|} \, L_{\frac{n-|l|}{2}}^{|l|}\!(r^2)
    \end{equation}
    with $n = 0,2,4,...$ and $l =-n,-n+2,...,+n$. The eigenenergy is $E_n = n+1$ with a degeneracy of $(n+1)$ for a single $n$ value.
    
    The axial part is the well-known problem of the harmonic oscillator in one dimension 
    \begin{equation}\label{eq:HOax}
        \left[ -\frac{1}{2}\partial_z^2 + \frac{1}{2}z^2 - E_k \right] \! Z_{k}(z) = 0 \,,
    \end{equation}
    that is solved by 
    \begin{equation}
        Z_{k}(z) = \sqrt{\frac{1}{\pi^{1/2} \, 2^k \, k!}} \, e^{-z^2/2} H_k(z)
    \end{equation}
    obtaining eigenenergies $E_k = k + \frac{1}{2}$ with $k = 0,1,2,...$.


    \bibliography{paper}

\begin{thebibliography}{27}%
\makeatletter
\providecommand \@ifxundefined [1]{%
 \@ifx{#1\undefined}
}%
\providecommand \@ifnum [1]{%
 \ifnum #1\expandafter \@firstoftwo
 \else \expandafter \@secondoftwo
 \fi
}%
\providecommand \@ifx [1]{%
 \ifx #1\expandafter \@firstoftwo
 \else \expandafter \@secondoftwo
 \fi
}%
\providecommand \natexlab [1]{#1}%
\providecommand \enquote  [1]{``#1''}%
\providecommand \bibnamefont  [1]{#1}%
\providecommand \bibfnamefont [1]{#1}%
\providecommand \citenamefont [1]{#1}%
\providecommand \href@noop [0]{\@secondoftwo}%
\providecommand \href [0]{\begingroup \@sanitize@url \@href}%
\providecommand \@href[1]{\@@startlink{#1}\@@href}%
\providecommand \@@href[1]{\endgroup#1\@@endlink}%
\providecommand \@sanitize@url [0]{\catcode `\\12\catcode `\$12\catcode
  `\&12\catcode `\#12\catcode `\^12\catcode `\_12\catcode `\%12\relax}%
\providecommand \@@startlink[1]{}%
\providecommand \@@endlink[0]{}%
\providecommand \url  [0]{\begingroup\@sanitize@url \@url }%
\providecommand \@url [1]{\endgroup\@href {#1}{\urlprefix }}%
\providecommand \urlprefix  [0]{URL }%
\providecommand \Eprint [0]{\href }%
\providecommand \doibase [0]{http://dx.doi.org/}%
\providecommand \selectlanguage [0]{\@gobble}%
\providecommand \bibinfo  [0]{\@secondoftwo}%
\providecommand \bibfield  [0]{\@secondoftwo}%
\providecommand \translation [1]{[#1]}%
\providecommand \BibitemOpen [0]{}%
\providecommand \bibitemStop [0]{}%
\providecommand \bibitemNoStop [0]{.\EOS\space}%
\providecommand \EOS [0]{\spacefactor3000\relax}%
\providecommand \BibitemShut  [1]{\csname bibitem#1\endcsname}%
\let\auto@bib@innerbib\@empty
\bibitem [{\citenamefont {Petrov}(2015)}]{Petrov2015}%
  \BibitemOpen
  \bibfield  {author} {\bibinfo {author} {\bibfnamefont {D.~S.}\ \bibnamefont
  {Petrov}},\ }\bibfield  {title} {\enquote {\bibinfo {title} {{Quantum
  Mechanical Stabilization of a Collapsing Bose-Bose Mixture}},}\ }\href
  {\doibase 10.1103/PhysRevLett.115.155302} {\bibfield  {journal} {\bibinfo
  {journal} {Phys. Rev. Lett.}\ }\textbf {\bibinfo {volume} {115}},\ \bibinfo
  {pages} {155302} (\bibinfo {year} {2015})}\BibitemShut {NoStop}%
\bibitem [{\citenamefont {Schmitt}\ \emph {et~al.}(2016)\citenamefont
  {Schmitt}, \citenamefont {Wenzel}, \citenamefont {B{\"{o}}ttcher},
  \citenamefont {Ferrier-Barbut},\ and\ \citenamefont {Pfau}}]{Schmitt2016}%
  \BibitemOpen
  \bibfield  {author} {\bibinfo {author} {\bibfnamefont {M.}~\bibnamefont
  {Schmitt}}, \bibinfo {author} {\bibfnamefont {M.}~\bibnamefont {Wenzel}},
  \bibinfo {author} {\bibfnamefont {F.}~\bibnamefont {B{\"{o}}ttcher}},
  \bibinfo {author} {\bibfnamefont {I.}~\bibnamefont {Ferrier-Barbut}}, \ and\
  \bibinfo {author} {\bibfnamefont {T.}~\bibnamefont {Pfau}},\ }\bibfield
  {title} {\enquote {\bibinfo {title} {{Self-bound droplets of a dilute
  magnetic quantum liquid}},}\ }\href {\doibase 10.1038/nature20126} {\bibfield
   {journal} {\bibinfo  {journal} {Nature}\ }\textbf {\bibinfo {volume}
  {539}},\ \bibinfo {pages} {259--262} (\bibinfo {year} {2016})}\BibitemShut
  {NoStop}%
\bibitem [{\citenamefont {Cabrera}\ \emph {et~al.}(2018)\citenamefont
  {Cabrera}, \citenamefont {Tanzi}, \citenamefont {Sanz}, \citenamefont
  {Naylor}, \citenamefont {Thomas}, \citenamefont {Cheiney},\ and\
  \citenamefont {Tarruell}}]{Cabrera2018}%
  \BibitemOpen
  \bibfield  {author} {\bibinfo {author} {\bibfnamefont {C.~R.}\ \bibnamefont
  {Cabrera}}, \bibinfo {author} {\bibfnamefont {L.}~\bibnamefont {Tanzi}},
  \bibinfo {author} {\bibfnamefont {J.}~\bibnamefont {Sanz}}, \bibinfo {author}
  {\bibfnamefont {B.}~\bibnamefont {Naylor}}, \bibinfo {author} {\bibfnamefont
  {P.}~\bibnamefont {Thomas}}, \bibinfo {author} {\bibfnamefont
  {P.}~\bibnamefont {Cheiney}}, \ and\ \bibinfo {author} {\bibfnamefont
  {L.}~\bibnamefont {Tarruell}},\ }\bibfield  {title} {\enquote {\bibinfo
  {title} {Quantum liquid droplets in a mixture of bose-einstein
  condensates},}\ }\href {\doibase 10.1126/science.aao5686} {\bibfield
  {journal} {\bibinfo  {journal} {Science}\ }\textbf {\bibinfo {volume}
  {359}},\ \bibinfo {pages} {301--304} (\bibinfo {year} {2018})}\BibitemShut
  {NoStop}%
\bibitem [{\citenamefont {Semeghini}\ \emph {et~al.}(2018)\citenamefont
  {Semeghini}, \citenamefont {Ferioli}, \citenamefont {Masi}, \citenamefont
  {Mazzinghi}, \citenamefont {Wolswijk}, \citenamefont {Minardi}, \citenamefont
  {Modugno}, \citenamefont {Modugno}, \citenamefont {Inguscio},\ and\
  \citenamefont {Fattori}}]{Semeghini2018}%
  \BibitemOpen
  \bibfield  {author} {\bibinfo {author} {\bibfnamefont {G.}~\bibnamefont
  {Semeghini}}, \bibinfo {author} {\bibfnamefont {G.}~\bibnamefont {Ferioli}},
  \bibinfo {author} {\bibfnamefont {L.}~\bibnamefont {Masi}}, \bibinfo {author}
  {\bibfnamefont {C.}~\bibnamefont {Mazzinghi}}, \bibinfo {author}
  {\bibfnamefont {L.}~\bibnamefont {Wolswijk}}, \bibinfo {author}
  {\bibfnamefont {F.}~\bibnamefont {Minardi}}, \bibinfo {author} {\bibfnamefont
  {M.}~\bibnamefont {Modugno}}, \bibinfo {author} {\bibfnamefont
  {G.}~\bibnamefont {Modugno}}, \bibinfo {author} {\bibfnamefont
  {M.}~\bibnamefont {Inguscio}}, \ and\ \bibinfo {author} {\bibfnamefont
  {M.}~\bibnamefont {Fattori}},\ }\bibfield  {title} {\enquote {\bibinfo
  {title} {Self-bound quantum droplets of atomic mixtures in free space},}\
  }\href {\doibase 10.1103/PhysRevLett.120.235301} {\bibfield  {journal}
  {\bibinfo  {journal} {Phys. Rev. Lett.}\ }\textbf {\bibinfo {volume} {120}},\
  \bibinfo {pages} {235301} (\bibinfo {year} {2018})}\BibitemShut {NoStop}%
\bibitem [{\citenamefont {Kadau}\ \emph {et~al.}(2016)\citenamefont {Kadau},
  \citenamefont {Schmitt}, \citenamefont {Wenzel}, \citenamefont {Wink},
  \citenamefont {Maier}, \citenamefont {Ferrier-Barbut},\ and\ \citenamefont
  {Pfau}}]{Kadau2016}%
  \BibitemOpen
  \bibfield  {author} {\bibinfo {author} {\bibfnamefont {H.}~\bibnamefont
  {Kadau}}, \bibinfo {author} {\bibfnamefont {M.}~\bibnamefont {Schmitt}},
  \bibinfo {author} {\bibfnamefont {M.}~\bibnamefont {Wenzel}}, \bibinfo
  {author} {\bibfnamefont {C.}~\bibnamefont {Wink}}, \bibinfo {author}
  {\bibfnamefont {T.}~\bibnamefont {Maier}}, \bibinfo {author} {\bibfnamefont
  {I.}~\bibnamefont {Ferrier-Barbut}}, \ and\ \bibinfo {author} {\bibfnamefont
  {T.}~\bibnamefont {Pfau}},\ }\bibfield  {title} {\enquote {\bibinfo {title}
  {{Observing the Rosensweig instability of a quantum ferrofluid}},}\ }\href
  {\doibase 10.1038/nature16485} {\bibfield  {journal} {\bibinfo  {journal}
  {Nature}\ }\textbf {\bibinfo {volume} {530}},\ \bibinfo {pages} {194--197}
  (\bibinfo {year} {2016})}\BibitemShut {NoStop}%
\bibitem [{\citenamefont {Ferrier-Barbut}\ \emph {et~al.}(2016)\citenamefont
  {Ferrier-Barbut}, \citenamefont {Kadau}, \citenamefont {Schmitt},
  \citenamefont {Wenzel},\ and\ \citenamefont {Pfau}}]{Ferrier-Barbut2016}%
  \BibitemOpen
  \bibfield  {author} {\bibinfo {author} {\bibfnamefont {I.}~\bibnamefont
  {Ferrier-Barbut}}, \bibinfo {author} {\bibfnamefont {H.}~\bibnamefont
  {Kadau}}, \bibinfo {author} {\bibfnamefont {M.}~\bibnamefont {Schmitt}},
  \bibinfo {author} {\bibfnamefont {M.}~\bibnamefont {Wenzel}}, \ and\ \bibinfo
  {author} {\bibfnamefont {T.}~\bibnamefont {Pfau}},\ }\bibfield  {title}
  {\enquote {\bibinfo {title} {{Observation of Quantum Droplets in a Strongly
  Dipolar Bose Gas}},}\ }\href {\doibase 10.1103/PhysRevLett.116.215301}
  {\bibfield  {journal} {\bibinfo  {journal} {Phys. Rev. Lett.}\ }\textbf
  {\bibinfo {volume} {116}},\ \bibinfo {pages} {215301} (\bibinfo {year}
  {2016})}\BibitemShut {NoStop}%
\bibitem [{\citenamefont {Chomaz}\ \emph {et~al.}(2016)\citenamefont {Chomaz},
  \citenamefont {Baier}, \citenamefont {Petter}, \citenamefont {Mark},
  \citenamefont {W{\"{a}}chtler}, \citenamefont {Santos},\ and\ \citenamefont
  {Ferlaino}}]{Chomaz2016a}%
  \BibitemOpen
  \bibfield  {author} {\bibinfo {author} {\bibfnamefont {L.}~\bibnamefont
  {Chomaz}}, \bibinfo {author} {\bibfnamefont {S.}~\bibnamefont {Baier}},
  \bibinfo {author} {\bibfnamefont {D.}~\bibnamefont {Petter}}, \bibinfo
  {author} {\bibfnamefont {M.~J.}\ \bibnamefont {Mark}}, \bibinfo {author}
  {\bibfnamefont {F.}~\bibnamefont {W{\"{a}}chtler}}, \bibinfo {author}
  {\bibfnamefont {L.}~\bibnamefont {Santos}}, \ and\ \bibinfo {author}
  {\bibfnamefont {F.}~\bibnamefont {Ferlaino}},\ }\bibfield  {title} {\enquote
  {\bibinfo {title} {{Quantum-fluctuation-driven crossover from a dilute
  Bose-Einstein condensate to a macro-droplet in a dipolar quantum fluid}},}\
  }\href {\doibase 10.1103/PhysRevX.6.041039} {\bibfield  {journal} {\bibinfo
  {journal} {Phys. Rev. X}\ }\textbf {\bibinfo {volume} {6}},\ \bibinfo {pages}
  {041039} (\bibinfo {year} {2016})}\BibitemShut {NoStop}%
\bibitem [{\citenamefont {Wenzel}\ \emph {et~al.}(2017)\citenamefont {Wenzel},
  \citenamefont {B\"ottcher}, \citenamefont {Langen}, \citenamefont
  {Ferrier-Barbut},\ and\ \citenamefont {Pfau}}]{Wenzel2017}%
  \BibitemOpen
  \bibfield  {author} {\bibinfo {author} {\bibfnamefont {M.}~\bibnamefont
  {Wenzel}}, \bibinfo {author} {\bibfnamefont {F.}~\bibnamefont {B\"ottcher}},
  \bibinfo {author} {\bibfnamefont {T.}~\bibnamefont {Langen}}, \bibinfo
  {author} {\bibfnamefont {I.}~\bibnamefont {Ferrier-Barbut}}, \ and\ \bibinfo
  {author} {\bibfnamefont {T.}~\bibnamefont {Pfau}},\ }\bibfield  {title}
  {\enquote {\bibinfo {title} {{Striped states in a many-body system of tilted
  dipoles}},}\ }\href {\doibase 10.1103/PhysRevA.96.053630} {\bibfield
  {journal} {\bibinfo  {journal} {Phys. Rev. A}\ }\textbf {\bibinfo {volume}
  {96}},\ \bibinfo {pages} {053630} (\bibinfo {year} {2017})}\BibitemShut
  {NoStop}%
\bibitem [{\citenamefont {W{\"{a}}chtler}\ and\ \citenamefont
  {Santos}(2016{\natexlab{a}})}]{Wachtler2016}%
  \BibitemOpen
  \bibfield  {author} {\bibinfo {author} {\bibfnamefont {F.}~\bibnamefont
  {W{\"{a}}chtler}}\ and\ \bibinfo {author} {\bibfnamefont {L.}~\bibnamefont
  {Santos}},\ }\bibfield  {title} {\enquote {\bibinfo {title} {{Quantum
  filaments in dipolar Bose-Einstein condensates}},}\ }\href {\doibase
  10.1103/PhysRevA.93.061603} {\bibfield  {journal} {\bibinfo  {journal} {Phys.
  Rev. A}\ }\textbf {\bibinfo {volume} {93}},\ \bibinfo {pages} {061603}
  (\bibinfo {year} {2016}{\natexlab{a}})}\BibitemShut {NoStop}%
\bibitem [{\citenamefont {Bisset}\ \emph {et~al.}(2016)\citenamefont {Bisset},
  \citenamefont {Wilson}, \citenamefont {Baillie},\ and\ \citenamefont
  {Blakie}}]{Bisset2016}%
  \BibitemOpen
  \bibfield  {author} {\bibinfo {author} {\bibfnamefont {R.~N.}\ \bibnamefont
  {Bisset}}, \bibinfo {author} {\bibfnamefont {R.~M.}\ \bibnamefont {Wilson}},
  \bibinfo {author} {\bibfnamefont {D.}~\bibnamefont {Baillie}}, \ and\
  \bibinfo {author} {\bibfnamefont {P.~B.}\ \bibnamefont {Blakie}},\ }\bibfield
   {title} {\enquote {\bibinfo {title} {{Ground-state phase diagram of a
  dipolar condensate with quantum fluctuations}},}\ }\href {\doibase
  10.1103/PhysRevA.94.033619} {\bibfield  {journal} {\bibinfo  {journal} {Phys.
  Rev. A}\ }\textbf {\bibinfo {volume} {94}},\ \bibinfo {pages} {033619}
  (\bibinfo {year} {2016})}\BibitemShut {NoStop}%
\bibitem [{\citenamefont {W{\"{a}}chtler}\ and\ \citenamefont
  {Santos}(2016{\natexlab{b}})}]{Wachtler2016a}%
  \BibitemOpen
  \bibfield  {author} {\bibinfo {author} {\bibfnamefont {F.}~\bibnamefont
  {W{\"{a}}chtler}}\ and\ \bibinfo {author} {\bibfnamefont {L.}~\bibnamefont
  {Santos}},\ }\bibfield  {title} {\enquote {\bibinfo {title} {{Ground-state
  properties and elementary excitations of quantum droplets in dipolar
  Bose-Einstein condensates}},}\ }\href {\doibase 10.1103/PhysRevA.94.043618}
  {\bibfield  {journal} {\bibinfo  {journal} {Phys. Rev. A}\ }\textbf {\bibinfo
  {volume} {94}},\ \bibinfo {pages} {043618} (\bibinfo {year}
  {2016}{\natexlab{b}})}\BibitemShut {NoStop}%
\bibitem [{\citenamefont {Baillie}\ \emph {et~al.}(2016)\citenamefont
  {Baillie}, \citenamefont {Wilson}, \citenamefont {Bisset},\ and\
  \citenamefont {Blakie}}]{Baillie2016}%
  \BibitemOpen
  \bibfield  {author} {\bibinfo {author} {\bibfnamefont {D.}~\bibnamefont
  {Baillie}}, \bibinfo {author} {\bibfnamefont {R.~M.}\ \bibnamefont {Wilson}},
  \bibinfo {author} {\bibfnamefont {R.~N.}\ \bibnamefont {Bisset}}, \ and\
  \bibinfo {author} {\bibfnamefont {P.~B.}\ \bibnamefont {Blakie}},\ }\bibfield
   {title} {\enquote {\bibinfo {title} {{Self-bound dipolar droplet: A
  localized matter wave in free space}},}\ }\href {\doibase
  10.1103/PhysRevA.94.021602} {\bibfield  {journal} {\bibinfo  {journal} {Phys.
  Rev. A}\ }\textbf {\bibinfo {volume} {94}},\ \bibinfo {pages} {021602}
  (\bibinfo {year} {2016})}\BibitemShut {NoStop}%
\bibitem [{\citenamefont {Baillie}\ \emph {et~al.}(2017)\citenamefont
  {Baillie}, \citenamefont {Wilson},\ and\ \citenamefont
  {Blakie}}]{Baillie2017}%
  \BibitemOpen
  \bibfield  {author} {\bibinfo {author} {\bibfnamefont {D.}~\bibnamefont
  {Baillie}}, \bibinfo {author} {\bibfnamefont {R.~M.}\ \bibnamefont {Wilson}},
  \ and\ \bibinfo {author} {\bibfnamefont {P.~B.}\ \bibnamefont {Blakie}},\
  }\bibfield  {title} {\enquote {\bibinfo {title} {Collective excitations of
  self-bound droplets of a dipolar quantum fluid},}\ }\href {\doibase
  10.1103/PhysRevLett.119.255302} {\bibfield  {journal} {\bibinfo  {journal}
  {Phys. Rev. Lett.}\ }\textbf {\bibinfo {volume} {119}},\ \bibinfo {pages}
  {255302} (\bibinfo {year} {2017})}\BibitemShut {NoStop}%
\bibitem [{\citenamefont {Saito}(2016)}]{Saito2016}%
  \BibitemOpen
  \bibfield  {author} {\bibinfo {author} {\bibfnamefont {H.}~\bibnamefont
  {Saito}},\ }\bibfield  {title} {\enquote {\bibinfo {title} {{Path-Integral
  Monte Carlo Study on a Droplet of a Dipolar Bose–Einstein Condensate
  Stabilized by Quantum Fluctuation}},}\ }\href {\doibase
  10.7566/JPSJ.85.053001} {\bibfield  {journal} {\bibinfo  {journal} {J. Phys.
  Soc. Japan}\ }\textbf {\bibinfo {volume} {85}},\ \bibinfo {pages} {053001}
  (\bibinfo {year} {2016})}\BibitemShut {NoStop}%
\bibitem [{\citenamefont {Macia}\ \emph {et~al.}(2016)\citenamefont {Macia},
  \citenamefont {S\'anchez-Baena}, \citenamefont {Boronat},\ and\ \citenamefont
  {Mazzanti}}]{Macia2016}%
  \BibitemOpen
  \bibfield  {author} {\bibinfo {author} {\bibfnamefont {A.}~\bibnamefont
  {Macia}}, \bibinfo {author} {\bibfnamefont {J.}~\bibnamefont
  {S\'anchez-Baena}}, \bibinfo {author} {\bibfnamefont {J.}~\bibnamefont
  {Boronat}}, \ and\ \bibinfo {author} {\bibfnamefont {F.}~\bibnamefont
  {Mazzanti}},\ }\bibfield  {title} {\enquote {\bibinfo {title} {{Droplets of
  Trapped Quantum Dipolar Bosons}},}\ }\href {\doibase
  10.1103/PhysRevLett.117.205301} {\bibfield  {journal} {\bibinfo  {journal}
  {Phys. Rev. Lett.}\ }\textbf {\bibinfo {volume} {117}},\ \bibinfo {pages}
  {205301} (\bibinfo {year} {2016})}\BibitemShut {NoStop}%
\bibitem [{\citenamefont {Cinti}\ and\ \citenamefont
  {Boninsegni}(2017)}]{Cinti2017}%
  \BibitemOpen
  \bibfield  {author} {\bibinfo {author} {\bibfnamefont {F.}~\bibnamefont
  {Cinti}}\ and\ \bibinfo {author} {\bibfnamefont {M.}~\bibnamefont
  {Boninsegni}},\ }\bibfield  {title} {\enquote {\bibinfo {title} {Classical
  and quantum filaments in the ground state of trapped dipolar bose gases},}\
  }\href {\doibase 10.1103/PhysRevA.96.013627} {\bibfield  {journal} {\bibinfo
  {journal} {Phys. Rev. A}\ }\textbf {\bibinfo {volume} {96}},\ \bibinfo
  {pages} {013627} (\bibinfo {year} {2017})}\BibitemShut {NoStop}%
\bibitem [{\citenamefont {Toennies}\ and\ \citenamefont
  {Vilesov}(2004)}]{Toennies2004}%
  \BibitemOpen
  \bibfield  {author} {\bibinfo {author} {\bibfnamefont {J.~P.}\ \bibnamefont
  {Toennies}}\ and\ \bibinfo {author} {\bibfnamefont {A.~F.}\ \bibnamefont
  {Vilesov}},\ }\bibfield  {title} {\enquote {\bibinfo {title} {{Superfluid
  helium droplets: A uniquely cold nanomatrix for molecules and molecular
  complexes}},}\ }\href {\doibase 10.1002/anie.200300611} {\bibfield  {journal}
  {\bibinfo  {journal} {Angew. Chemie}\ }\textbf {\bibinfo {volume} {43}},\
  \bibinfo {pages} {2622--2648} (\bibinfo {year} {2004})}\BibitemShut {NoStop}%
\bibitem [{\citenamefont {Lahaye}\ \emph {et~al.}(2009)\citenamefont {Lahaye},
  \citenamefont {Menotti}, \citenamefont {Santos}, \citenamefont {Lewenstein},\
  and\ \citenamefont {Pfau}}]{Lahaye2009}%
  \BibitemOpen
  \bibfield  {author} {\bibinfo {author} {\bibfnamefont {T.}~\bibnamefont
  {Lahaye}}, \bibinfo {author} {\bibfnamefont {C.}~\bibnamefont {Menotti}},
  \bibinfo {author} {\bibfnamefont {L.}~\bibnamefont {Santos}}, \bibinfo
  {author} {\bibfnamefont {M.}~\bibnamefont {Lewenstein}}, \ and\ \bibinfo
  {author} {\bibfnamefont {T.}~\bibnamefont {Pfau}},\ }\bibfield  {title}
  {\enquote {\bibinfo {title} {{The physics of dipolar bosonic quantum
  gases}},}\ }\href {\doibase 10.1088/0034-4885/72/12/126401} {\bibfield
  {journal} {\bibinfo  {journal} {Rep. Prog. Phys.}\ }\textbf {\bibinfo
  {volume} {72}},\ \bibinfo {pages} {126401} (\bibinfo {year}
  {2009})}\BibitemShut {NoStop}%
\bibitem [{\citenamefont {Lu}\ \emph {et~al.}(2012)\citenamefont {Lu},
  \citenamefont {Burdick},\ and\ \citenamefont {Lev}}]{Lu2012}%
  \BibitemOpen
  \bibfield  {author} {\bibinfo {author} {\bibfnamefont {M.}~\bibnamefont
  {Lu}}, \bibinfo {author} {\bibfnamefont {N.~Q.}\ \bibnamefont {Burdick}}, \
  and\ \bibinfo {author} {\bibfnamefont {B.~L.}\ \bibnamefont {Lev}},\
  }\bibfield  {title} {\enquote {\bibinfo {title} {Quantum degenerate dipolar
  fermi gas},}\ }\href {\doibase 10.1103/PhysRevLett.108.215301} {\bibfield
  {journal} {\bibinfo  {journal} {Phys. Rev. Lett.}\ }\textbf {\bibinfo
  {volume} {108}},\ \bibinfo {pages} {215301} (\bibinfo {year}
  {2012})}\BibitemShut {NoStop}%
\bibitem [{\citenamefont {Sch{\"{u}}tzhold}\ \emph {et~al.}(2006)\citenamefont
  {Sch{\"{u}}tzhold}, \citenamefont {Uhlmann}, \citenamefont {Xu},\ and\
  \citenamefont {Fischer}}]{Schutzhold2006}%
  \BibitemOpen
  \bibfield  {author} {\bibinfo {author} {\bibfnamefont {R.}~\bibnamefont
  {Sch{\"{u}}tzhold}}, \bibinfo {author} {\bibfnamefont {M.}~\bibnamefont
  {Uhlmann}}, \bibinfo {author} {\bibfnamefont {Y.}~\bibnamefont {Xu}}, \ and\
  \bibinfo {author} {\bibfnamefont {U.~R.}\ \bibnamefont {Fischer}},\
  }\bibfield  {title} {\enquote {\bibinfo {title} {{Mean-Field Expansion in
  Bose–Einstein Condensates With Finite-Range Interactions}},}\ }\href
  {\doibase 10.1142/S0217979206035631} {\bibfield  {journal} {\bibinfo
  {journal} {Int. J. Mod. Phys. B}\ }\textbf {\bibinfo {volume} {20}},\
  \bibinfo {pages} {3555} (\bibinfo {year} {2006})}\BibitemShut {NoStop}%
\bibitem [{\citenamefont {Lima}\ and\ \citenamefont
  {Pelster}(2011)}]{Lima2011}%
  \BibitemOpen
  \bibfield  {author} {\bibinfo {author} {\bibfnamefont {A.~R.~P.}\
  \bibnamefont {Lima}}\ and\ \bibinfo {author} {\bibfnamefont {A.}~\bibnamefont
  {Pelster}},\ }\bibfield  {title} {\enquote {\bibinfo {title} {{Quantum
  fluctuations in dipolar Bose gases}},}\ }\href {\doibase
  10.1103/PhysRevA.84.041604} {\bibfield  {journal} {\bibinfo  {journal} {Phys.
  Rev. A}\ }\textbf {\bibinfo {volume} {84}},\ \bibinfo {pages} {041604}
  (\bibinfo {year} {2011})}\BibitemShut {NoStop}%
\bibitem [{Note1()}]{Note1}%
  \BibitemOpen
  \bibinfo {note} {We assume equal masses $m_\protect \mathrm {f} \approx
  m_\protect \mathrm {b}$ and therefore use $m_\protect \mathrm {f}$ instead of
  introducing the reduced mass.}\BibitemShut {Stop}%
\bibitem [{\citenamefont {Ferrier-Barbut}\ \emph {et~al.}(2018)\citenamefont
  {Ferrier-Barbut}, \citenamefont {Wenzel}, \citenamefont {B{\"{o}}ttcher},
  \citenamefont {Langen}, \citenamefont {Isoard}, \citenamefont {Stringari},\
  and\ \citenamefont {Pfau}}]{Ferrier-Barbut2018}%
  \BibitemOpen
  \bibfield  {author} {\bibinfo {author} {\bibfnamefont {I.}~\bibnamefont
  {Ferrier-Barbut}}, \bibinfo {author} {\bibfnamefont {M.}~\bibnamefont
  {Wenzel}}, \bibinfo {author} {\bibfnamefont {F.}~\bibnamefont
  {B{\"{o}}ttcher}}, \bibinfo {author} {\bibfnamefont {T.}~\bibnamefont
  {Langen}}, \bibinfo {author} {\bibfnamefont {M.}~\bibnamefont {Isoard}},
  \bibinfo {author} {\bibfnamefont {S.}~\bibnamefont {Stringari}}, \ and\
  \bibinfo {author} {\bibfnamefont {T.}~\bibnamefont {Pfau}},\ }\bibfield
  {title} {\enquote {\bibinfo {title} {{Scissors Mode of Dipolar Quantum
  Droplets of Dysprosium Atoms}},}\ }\href {\doibase
  10.1103/PhysRevLett.120.160402} {\bibfield  {journal} {\bibinfo  {journal}
  {Phys. Rev. Lett.}\ }\textbf {\bibinfo {volume} {120}},\ \bibinfo {pages}
  {160402} (\bibinfo {year} {2018})}\BibitemShut {NoStop}%
\bibitem [{\citenamefont {Maier}\ \emph {et~al.}(2015)\citenamefont {Maier},
  \citenamefont {Kadau}, \citenamefont {Schmitt}, \citenamefont {Wenzel},
  \citenamefont {Ferrier-Barbut}, \citenamefont {Pfau}, \citenamefont {Frisch},
  \citenamefont {Baier}, \citenamefont {Aikawa}, \citenamefont {Chomaz},
  \citenamefont {Mark}, \citenamefont {Ferlaino}, \citenamefont {Makrides},
  \citenamefont {Tiesinga}, \citenamefont {Petrov},\ and\ \citenamefont
  {Kotochigova}}]{Maier2015a}%
  \BibitemOpen
  \bibfield  {author} {\bibinfo {author} {\bibfnamefont {T.}~\bibnamefont
  {Maier}}, \bibinfo {author} {\bibfnamefont {H.}~\bibnamefont {Kadau}},
  \bibinfo {author} {\bibfnamefont {M.}~\bibnamefont {Schmitt}}, \bibinfo
  {author} {\bibfnamefont {M.}~\bibnamefont {Wenzel}}, \bibinfo {author}
  {\bibfnamefont {I.}~\bibnamefont {Ferrier-Barbut}}, \bibinfo {author}
  {\bibfnamefont {T.}~\bibnamefont {Pfau}}, \bibinfo {author} {\bibfnamefont
  {A.}~\bibnamefont {Frisch}}, \bibinfo {author} {\bibfnamefont
  {S.}~\bibnamefont {Baier}}, \bibinfo {author} {\bibfnamefont
  {K.}~\bibnamefont {Aikawa}}, \bibinfo {author} {\bibfnamefont
  {L.}~\bibnamefont {Chomaz}}, \bibinfo {author} {\bibfnamefont {M.~J.}\
  \bibnamefont {Mark}}, \bibinfo {author} {\bibfnamefont {F.}~\bibnamefont
  {Ferlaino}}, \bibinfo {author} {\bibfnamefont {C.}~\bibnamefont {Makrides}},
  \bibinfo {author} {\bibfnamefont {E.}~\bibnamefont {Tiesinga}}, \bibinfo
  {author} {\bibfnamefont {A.}~\bibnamefont {Petrov}}, \ and\ \bibinfo {author}
  {\bibfnamefont {S.}~\bibnamefont {Kotochigova}},\ }\bibfield  {title}
  {\enquote {\bibinfo {title} {{Emergence of Chaotic Scattering in Ultracold Er
  and Dy}},}\ }\href {\doibase 10.1103/PhysRevX.5.041029} {\bibfield  {journal}
  {\bibinfo  {journal} {Phys. Rev. X}\ }\textbf {\bibinfo {volume} {5}},\
  \bibinfo {pages} {041029} (\bibinfo {year} {2015})}\BibitemShut {NoStop}%
\bibitem [{\citenamefont {Bergschneider}\ \emph {et~al.}(2018)\citenamefont
  {Bergschneider}, \citenamefont {Klinkhamer}, \citenamefont {Becher},
  \citenamefont {Klemt}, \citenamefont {Z\"urn}, \citenamefont {Preiss},\ and\
  \citenamefont {Jochim}}]{Bergschneider2018}%
  \BibitemOpen
  \bibfield  {author} {\bibinfo {author} {\bibfnamefont {A.}~\bibnamefont
  {Bergschneider}}, \bibinfo {author} {\bibfnamefont {V.~M.}\ \bibnamefont
  {Klinkhamer}}, \bibinfo {author} {\bibfnamefont {J.~H.}\ \bibnamefont
  {Becher}}, \bibinfo {author} {\bibfnamefont {R.}~\bibnamefont {Klemt}},
  \bibinfo {author} {\bibfnamefont {G.}~\bibnamefont {Z\"urn}}, \bibinfo
  {author} {\bibfnamefont {P.~M.}\ \bibnamefont {Preiss}}, \ and\ \bibinfo
  {author} {\bibfnamefont {S.}~\bibnamefont {Jochim}},\ }\bibfield  {title}
  {\enquote {\bibinfo {title} {Spin-resolved single-atom imaging of
  $^{6}\mathrm{Li}$ in free space},}\ }\href {\doibase
  10.1103/PhysRevA.97.063613} {\bibfield  {journal} {\bibinfo  {journal} {Phys.
  Rev. A}\ }\textbf {\bibinfo {volume} {97}},\ \bibinfo {pages} {063613}
  (\bibinfo {year} {2018})}\BibitemShut {NoStop}%
\bibitem [{\citenamefont {Boudjemâa}(2017)}]{Boudjemaa2017}%
  \BibitemOpen
  \bibfield  {author} {\bibinfo {author} {\bibfnamefont {A.}~\bibnamefont
  {Boudjemâa}},\ }\bibfield  {title} {\enquote {\bibinfo {title} {{Quantum
  dilute droplets of dipolar bosons at finite temperature}},}\ }\href {\doibase
  10.1016/j.aop.2017.03.020} {\bibfield  {journal} {\bibinfo  {journal} {Annals
  of Physics}\ }\textbf {\bibinfo {volume} {381}},\ \bibinfo {pages} {68}
  (\bibinfo {year} {2017})}\BibitemShut {NoStop}%
\bibitem [{\citenamefont {Aybar}\ and\ \citenamefont
  {Oktel}(2018)}]{Aybar2018}%
  \BibitemOpen
  \bibfield  {author} {\bibinfo {author} {\bibfnamefont {E.}~\bibnamefont
  {Aybar}}\ and\ \bibinfo {author} {\bibfnamefont {M.~O.}\ \bibnamefont
  {Oktel}},\ }\bibfield  {title} {\enquote {\bibinfo {title} {{Temperature
  Dependent Density Profiles of Dipolar Droplets}},}\ }\href
  {http://arxiv.org/abs/1805.06261} {\bibfield  {journal} {\bibinfo  {journal}
  {arXiv:1805.06261}\ } (\bibinfo {year} {2018})}\BibitemShut {NoStop}%
\end{thebibliography}%

\end{document}